\documentclass[12pt,preprint]{aastex}  % manuscript
\usepackage{psfig}

\newcommand {\ea} {{\it et~al.}}
\newcommand {\be} {\begin{equation}}
\newcommand {\ee} {\end{equation}}

\shorttitle{X-rays from IC 2560}

\shortauthors{madejski \ea}

\begin{document}

\title{X-ray Emission from Megamaser Galaxy IC 2560}

\author{G. Madejski\altaffilmark{1,2},
C. Done\altaffilmark{3},
P. T. \.{Z}ycki\altaffilmark{4}, and
L. Greenhill\altaffilmark{2,5}
}

\altaffiltext{1}{Stanford Linear Accelerator Center, 2575 Sand Hill Road,
Menlo Park, CA 94025, USA;  \tt{madejski@stanford.edu}}
\altaffiltext{2}{Kavli Institute for Particle Astrophysics and Cosmology, 
Stanford, CA 94305} 
\altaffiltext{3}{Dept. of Physics, Durham University, Durham, UK}
\altaffiltext{4}{Nicolaus Copernicus Astronomical Center, Warsaw, Poland}
\altaffiltext{5}{Harvard-Smithsonian Center for Astrophysics, 
Cambridge, MA 02138, USA}

\begin{abstract}

Observation of the H$_2$O megamaser galaxy IC~2560 with the 
Chandra Observatory reveals a complex spectrum composed of 
soft X-ray emission due to multi-temperature thermal plasma, 
and a hard 
continuum with strong emission lines. The continuum is most 
likely a Compton reflection (reprocessing) of primary emission that 
is completely absorbed at least up to 7 keV. The lines can be 
identified with fluorescence from Si, S and Fe in the lowest 
ionization stages. The equivalent widths of the Si and S lines 
are broadly compatible with those anticipated for reprocessing 
by optically thick cold plasma of Solar abundances, while the 
large equivalent width of the Fe line requires some overabundance 
of iron.  A contribution to the line from a transmitted component 
cannot be ruled out, but the limits on the 
strength of the Compton shoulder make it less likely.  
From the bolometric luminosity of the 
nuclear region, we infer that the source radiates at 1 - 10\% of its 
Eddington luminosity,   for an adopted central mass 
of $3\times10^6$ M$_\odot$.  The overall spectrum is 
consistent with the hypotheses that the central 
engines powering the detected megamsers in accretion disks are 
obscured from direct view by the associated accretion
disk material itself, and that there is a correlation between 
the occurrence of megamaser emission and Compton-thick 
absorption columns.  For the 11 known galaxies with both
column density measurements and maser emission believed to arise from
accretion disks, eight AGN are Compton thick.

\end{abstract}


\section{Introduction}

Position-velocity resolved interferometric mapping of water vapor 
maser emission ($\nu_{rest}=22235.08$~MHz) in active galactic nuclei
(AGN) is a powerful tool for the study of gas structures and dynamics in 
regions less than of order 1 pc from massive central engines. For the 
cases in which the megamaser emission can be identified with an 
accretion disk rather than with the wind- or jet-like outflow,  such 
mapping may allow the determination central engine (and presumably 
black hole) mass, Eddington luminosity, and disk geometry (e.g., warping).  
The three most notable examples are: the geometrically thin 
accretion disk of NGC~4258 (Myioshi et al.\ 1995; Greenhill et al.\ 1995) 
which obeys a  Keplerian rotation law to $\ll 1\%$; the apparently massive 
accretion disk within the NGC~1068 AGN (Greenhill et al. 1997; 
Kumar 1999; Lodato \& Bertin 2003); and the accretion disk within the 
nucleus of the Circinus Galaxy (Greenhill et al. \ 2003), and the associated 
wide angle wind.

Water megamaser and nuclear activity are linked by the maser pump 
process which in the case of accretion 
disks, is plausibly driven by collisions in molecular gas heated by X-ray 
irradiation (Neufeld, Maloney, \& Conger 1994).  Nonetheless, detectable 
megamaser emission is relatively rare.
This is probably a consequence of intrinsic luminosity and geometry.  
In general, maser emission is amplified most along the directions in 
which there is the most material with overlapping line-of-sight 
velocities, and as a result the emission is beamed narrowly along 
tangent planes for any individual accretion disk. Association with 
type-2 AGN is a consequence. 

Average detection rates among type-2 AGN are on the order of
10\% (e. g. Braatz et al. 2004), 
depending on galaxy distance and instrument sensitivity.   Only
$\sim 60$ megamaser systems are known (Kondratko et al. 2005b, c;  
Henkel et al. 2005 and references therein) and evidence from 
spectra and Very Long Baseline Interferometer (VLBI) maps associating the 
emission with accretion disks is available for about a third 
(e.g., Kondratko, Greenhill, \& Moran 2005a).   Nonetheless, 
the contribution to
understanding of AGN by broadband studies of megamaser galaxies lies 
in the qualitatively greater level of detail that may be achieved, 
specifically because the structure and dynamics of the molecular 
gas can be so well constrained.  

Though the central engines associated with megamasers are likely 
to be obscured in the optical, UV, and soft X-ray bands -- at least 
for the absorbing columns
corresponding to less than a few Thomson optical depths -- X-ray
spectroscopy can provide a good estimate of the  absorbing column, and
thus the true luminosity of the central source.   For other, more heavily
obscured sources, the details of the profile of the Fe K$\alpha$
line can often provide additional constraints (e.g., Levenson et al.
2002).    One good candidate for detailed measurements is the Seyfert 2
galaxy IC~2560.  It contains an H$_2$O megamaser (Braatz, Wilson,\& 
Henkel 1996), with a peak flux density of up to $\sim 0.4$ Jy.
This relatively nearby (26 Mpc; $v_{rec} = 2876$ km s$^{-1}$) source has 
been partially resolved with the VLBI (Ishihara et al. 2001) 
and a central black hole mass of $\sim 3 \times 10^6$ M$_\odot$ inferred, 
which is relatively low but comparable to the mass of the black hole in 
our own Milky Way (Ghez et al. 2005;  Sch\"odel et al. 2003) and 
Circinus Galaxy (Greenhill et al. 2003).  

Prior to the observation described in Section 2, this object has been observed 
with Chandra in 2000 October, via a short ($\sim 10$
ks) observation, reported in Iwasawa, Maloney, \& Fabian (2002), 
and even before that, with Asca (Ishihara et al. 2001).  The data 
reported here are based on a longer, 50 ks pointing obtained 
in 2004 February.

\section{Observations, data reduction, and spectral fitting}

A 50 ks Chandra observation of IC~2560 was conducted on 2004 February
16-17. The processed data were reduced using the XSELECT tool, with an
independent analysis using CIAO.  Data were collected with ACIS-S,
with the source on chip S3.  The intensity of unrejected background
was constant and relatively low (less than 2 ct s$^{-1}$ in the entire
ACIS-S3 chip), showing no sign of any flares, so removal of 
any additional data segments was not necessary. The image consists mainly of
the point source, but there is a hint of some extended component. The
spectrum of the extended component is under investigation.  To measure
the spectrum of the nuclear source, we extracted the data from a
region $2''$ ($\sim 250$ pc) in radius, and determined the background
from an annulus with inner radius $30''$ and outer radius $90''$ centered 
on the source.  This corresponds to roughly 3.8 kpc and 11.3 kpc at the 
source's distance, and while might also include some of the diffuse emission
from the galaxy itself, in the quite compact source region chosen here,
the contribution of such diffuse emission (or instrumental background) 
was negligible, less than 0.1\% of the source count rate.
Total net count rate in the 0.5 - 10.0 keV range was 0.025 ct
s$^{-1}$.  The spectrum was subsequently rebinned to allow at least 20
counts in a new spectral bin.  We note that there is no variability in
the source flux during the Chandra observation.  Furthermore, the source 
shows the same count rate -- as well as flux, for the same assumed models --
as reported in Iwasawa et al. (2002).  We prepared
the ACIS S3 redistribution matrices and the effective area files as
appropriate for this observation.  Since the data were taken
relatively recently, the CTI correction has been applied in the course
of the pipeline processing.

\begin{deluxetable}{lccccccc}
\rotate

\label{tab:spectralmodels}
\tablewidth{0pc}
\tablecolumns{8}

\tablecaption{Results of data modeling without photo-ionization} 
\tablehead{
\colhead { model\tablenotemark{a}}
     & $kT_1$\tablenotemark{b} (keV) & $A_1$\tablenotemark{b}  &
       $kT_2$\tablenotemark{b} (keV) & $A_2$\tablenotemark{b} &
       $\Gamma$\tablenotemark{c} & EW (keV)\tablenotemark{d} & 
$\chi^2$/dof }
\startdata

1 mk+po+ga   &   $0.53^{+0.13}_{-0.09}$  & $0.019^{-0.014}_{-0.013}$
& -- & -- & $-0.3\pm 0.3$ & $2.7 \pm 0.5$ & 89.6/49 \\
2 mk+rf+ga    & $0.58^{+0.10}_{-0.14}$ & $0.023_{-0.14}^{+0.023}$ & --
& -- & $2.24^{+0.45}_{-0.39}$ & $2.8_{-0.6}^{+0.07}$ & 90.2/49 \\
3 2mk+rf+ga & $0.65^{+0.10}_{-0.06}$  & $1^e$ & $0.14^{+0.05}_{-0.05e}$ &
$1^e$ & $2.8\pm 0.2$ & $3.6^{+0.8}_{-0.6}$ & 83.7/48 \\
4 2mk+rf+3ga$^f$ & $0.65_{-0.06}^{+0.11}$  & $1^e$ & 
$0.14_{-0.06g}^{+0.04}$
& $1^e$ & $2.7_{-0.3}^{+0.2}$ & $3.4\pm 0.7$ & 71.1/44 \\
5 ab+2mk+rf+3ga+ds$^h$ & $0.58^{+0.07}_{-0.13}$ & $1^e$ &
$0.08_{-0f}^{+0.02}$  & $1^e$ & $2.2\pm 0.5$ & $2.7_{-0.6}^{+0.4}$ & 61/42 
\\

\enddata

\tablenotetext{a}{Those describe spectral forms incorporated into the 
model:
mk -- {\tt mekal}, po -- power law, \\
ga -- Gaussian line,
rf -- Compton reflection ({\tt pexrav}), 3ga -- three Gaussian lines, 
ds -- Compton \\ 
down-scattered line shoulder, 
ab -- additional absorption at redshift $z=0.01$ of host 
galaxy}
\tablenotetext{b}{{\tt mekal} model plasma temperature and metal 
abundances}
\tablenotetext{c}{Power law photon spectral index, $N(E) \propto 
E^{-\Gamma}$ }
\tablenotetext{d}{EW is the equivalent width of the Fe 6.4 keV line }
\tablenotetext{e}{Parameter fixed; the abundances are not well
constrained ($0.2 < A < 1$) }
\tablenotetext{f}{Three Gaussian lines are the Fe-K, Si-K, and S-K
lines originating in neutral material, \\
as described in the text}
\tablenotetext{g}{Parameter uncertainty pegged at lower limit}
\tablenotetext{h}{Host galaxy $N_H=0.66_{-0.24}^{+0.08}\times 10^{22}\,
{\rm cm}^{-2}$; and upper limit to the downscattered/observed \\
line intensity of $\sim 0.15$}

\end{deluxetable}

\begin{deluxetable}{lcccccc}
\rotate
\label{tab:xstar}
\tablewidth{0pc}
\tablecolumns{7}
\tablecaption{Results of data modeling with photo-ionization models}
\tablehead{
\colhead { model\tablenotemark{a}}
     & $\log(\xi_1)$\tablenotemark{b} & $\log(\xi_2)$\tablenotemark{b}  &
       $k T$\tablenotemark{c} (keV) &
       $\Gamma$\tablenotemark{d} & EW (keV)\tablenotemark{e} & 
$\chi^2$/dof }
\startdata

1 xstar\tablenotemark{f}+rf+3ga   &   1.6  & -- & -- &
   $3.11^{+0.14}_{-0.15}$ &
   $4.6^{+1.0}_{-0.8}$ & 152/49 \\
2 xstar\tablenotemark{f}+po+rf+3ga  & 1.6 & -- & -- &
   $2.2^{+0.3}_{-0.4}$ &
     3.5 & 106/48 \\
3 xstar\tablenotemark{f}+mk+rf+3ga & 3.0 & --  & $0.23^{+0.03}_{-0.01}$ &
    $1.5^{+0.5}_{-0.4}$ & $2.7^{+0.3}_{-0.8}$ &
      89.4/47  \\
4 2xstar\tablenotemark{f}+rf+3ga &  3.0  & 1.4 & -- & $1.5^{+0.5}_{-0.4}$ 
&
     $2.4\pm 0.5$ & 96.4/47 \\
5 2xstar\tablenotemark{f}+mk+rf+3ga & 3.0 & $1.6 \pm 0.2$  & $0.19 \pm 
0.03$ &
   $1.6 \pm 0.4$ &
    $2.40 \pm 0.45$ & 58.1/45
\\

\enddata

\tablenotetext{a}{Spectral components: xstar - {\sc XSTAR} 
photo-ionization
model;  other components as in Table~1}
\tablenotetext{b}{Log of ionization parameter in {\sc XSTAR} model}
\tablenotetext{c}{Plasma temperature in {\tt mekal} model}
\tablenotetext{d}{Power law photon spectral index, $N(E) \propto 
E^{-\Gamma}$ }
\tablenotetext{e}{EW is the equivalent width of the Fe 6.4 keV line }
\tablenotetext{f}{Model parameterized by ionization parameter $\xi$
and H column density, $N_H$. Fits show no dependence on $N_H$,
uncertainties on $\log(\xi)$ are usually smaller than model grid
spacing ($\Delta\log(\xi) = 0.2$), and are not shown}

\end{deluxetable}

The background-subtracted data are well-represented by a spectrum with
three general components: a very hard continuum, a soft component with
some soft X-ray emission lines, and an emission line complex between 6
and 7 keV, presumably due to Fe K-shell transitions. This is illustrated
in Figure 1 where we assumed as a model a simple power law, absorbed
by the Galactic column of $6.5 \times 10^{20}$ cm$^{-2}$.  
The residuals clearly show the features above.  This spectrum is
similar to that reported by Iwasawa et al.\ (2002), but the longer
exposure clearly reveals more details, with better resolved spectral
features. A somewhat more complex and realistic model
including a power law, collisionally ionized
plasma (described as a {\tt mekal} XSPEC model;  see below) and Gaussian
line, respectively, for the hard continuum, soft emission, and the Fe
line, gives an energy power law index $\alpha = -1.3\pm 0.3$
(Table~1, Model 1).  This is extremely hard for
the intrinsic X-ray spectrum of an AGN, but since IC~2560
is a megamaser source, the AGN might be
obscured by a large column of absorbing material and the continuum
measured here is probably not the primary spectrum.  Specifically,
the primary continuum might well be a power law, but we anticipate that it
is heavily absorbed rather than particularly hard (as compared
to unobscured Seyferts), and it may be accompanied by a Compton
reflection component, which is likely to be present when the primary
source photons are  reprocessed by circumnuclear material. If the
absorption is particularly severe --
$\sim 10^{25}$ cm$^{-2}$ or more for Solar abundances -- the very hard
spectrum measured above $\sim 2.5$ keV could be just that due to the
reflected component, although some additional flux might arise via
transmission through the absorber.

\begin{figure}
\includegraphics[angle=270,scale=0.5]{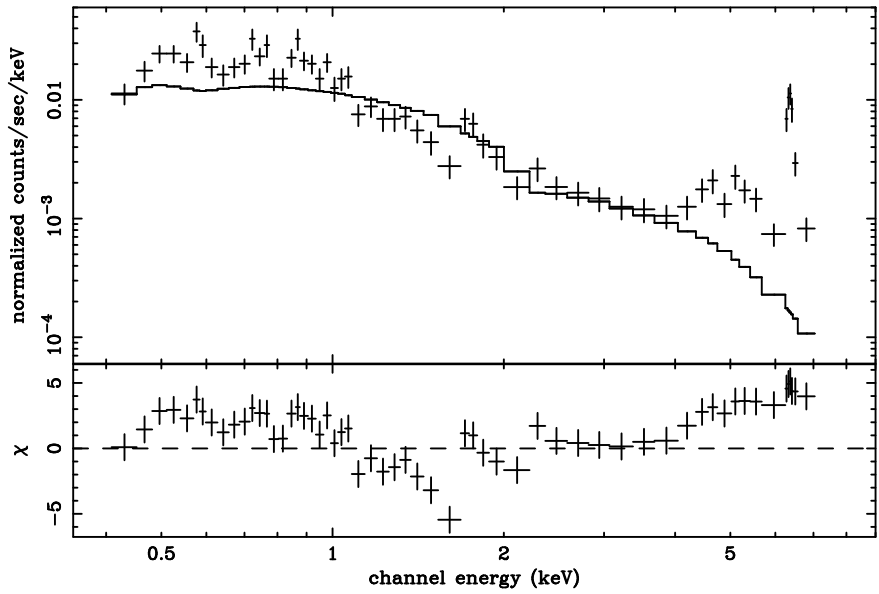}
\caption{Chandra count spectrum for IC~2560, fitted to a simple
power law model, absorbed at low energies by the Galactic absorption of
$6.5 \times 10^{20}$ cm$^{-2}$.  The residuals to the fit
(plotted as the contribution to $\chi^{2}$) show
strong residual features, modeled and discussed in Sec. 2.}
\end{figure}

Below, we describe the hard continuum as pure Compton reflection.
When we replace the power law component representing the
primary continuum with a reflection component at a
fixed $60^\circ$ inclination (using the XSPEC model {\tt pexrav};  see
Magdziarz \&  Zdziarski 1995), we obtain $\alpha=1.3^{+0.3}_{-0.4}$,
which corresponds to a more reasonable intrinsic spectrum
of the illuminating photons.  The choice of a pure reflection model is
also supported by the detection of the very strong,
intrinsically narrow Fe K line (equivalent width $\sim 2.5$ keV,
depending on the adopted continuum model, and $\sigma < 40$~eV) at an
energy $6.40^{+0.1}_{-0.2}$~keV, which is consistent with neutral
material and in  agreement with Iwasawa et al.\ (2002). The soft
component below 1~keV is fairly well fit by a single temperature,
collisionally ionized plasma with $kT=0.5\pm 0.1$~keV, but with
unphysically low abundances $A=0.02\pm 0.01$ (Table~1, Model 2). 

The strong iron line
should also be accompanied by a noticeable Compton downscattered shoulder,
forming a continuum centered at a  rest frame energy of $\sim 6.3$~keV,
with intrinsic width of $\sim 0.14$~keV (Illarionov et al.\ 1979). The
strength of this continuum relative to the unscattered narrow line core
depends on details of the reflecting material (Matt 2002;  see Fig. 2).
This component has been seen in other reflection-dominated AGN such as
Circinus Galaxy (Molendi, Bianchi, \& Matt 2003), NGC~4945  (Done et al. 2003),
and NGC~1068 (Matt et al. 2004). We incorporated the shape of this
downscattered continuum (Illarionov et al.\ 1979) as a local model in
{\sc xspec}, assuming a temperature of 1~eV for the reflecting material, and 
include it in the spectral fit in Table 1, Model 5.  
While this downscattered continuum 
is not significantly detected in our data, the 90\%
confidence upper limit on its strength is $\sim 15$\% of that
of the narrow line core. The low value indicates
super-Solar iron abundance, with the observed flux arising
from a reflector viewed at a high inclination (Matt 2002),
as may be anticipated for reflection from the far side of a
flattened structure with fairly constant ratio of height to
radius.

In addition to the 6.4 keV line, there is probably a marginally resolved 
line at 6.7 keV as well,  although it is not highly significant (and thus not 
included in any fits in Table~1):  adding three parameters improves 
the best-fit $\chi^2$ only by 5.   Close inspection of the unbinned 
data does not reveal whether the line is intrinsically broad or
divisible into sub-components, but the latter is a more likely possibility.  
If real, the 6.7 keV emission indicates very hot or highly ionized material, 
in contrast to the neutral material responsible for the 1.75 (Si) and 
2.32~keV (S) lines discussed below, and the Compton reflecting material, 
which has a lower inferred ionization parameter. Presence of 6.7~keV 
emission provides marginal evidence of a multi-temperature plasma in 
close proximity to the central engine.  

Rather than a very low abundance single-temperature plasma, the
soft spectrum may be well fit by a model comprising nearly
Solar abundance material, but also including second plasma component
with $kT \sim 0.1$ keV.  The new component contributes the additional continuum
necessary to dilute model line strengths and match the
observations with more typical abundances (Table~1 as Model 3).  These
two temperature components are probably simply an approximation to the
multi-temperature hot gas which is expected in starburst regions (e.g.\
Strickland \& Stevens 2000).  It should not be confused with the
multi-temperature plasma inferred from the Fe line emission, which
originates on much smaller scales. We note here that even though a
2-temperature model is consistent  with Solar abundances, it constrains
them poorly:  spectral fits -- where we require the same abundance of
both {\tt mekal} components --  allow abundances ranging from 0.2 to 1.
We note here that it is possible  to fit the spectrum with a combination
of thermal plasma and photo-ionized  components, but at least some
contribution from thermal  plasma is required;  we discuss this more
extensively below.   However, there is {\em no\/} combination of hot gas 
components which can account for the major residuals in the spectral fit
at $\sim 1.8$ and $2.3$~keV. This is because these features are at the energies
expected from {\em neutral\/} Si and S, so cannot be produced by hot
plasma emission. The obvious cool plasma component is the one which
produces the reflected continuum. For Solar abundances this should
produce an equivalent width of $\sim 250$ and $170$ eV in neutral Si and
S, respectively (Matt, Fabian, \& Reynolds 1997) to accompany the Fe line
of equivalent width 0.8--1.6~keV (depending on atomic data and assumed
geometry: George \& Fabian 1991; \.{Z}ycki \& Czerny 1994, Matt, Brandt,
\& Fabian 1996). Adding in two additional Gaussian lines gives Si line
emission with equivalent width  of $130_{-80}^{+90}$~eV at $1.75\pm
0.03$~keV and S of $200\pm 140$ eV at $2.32\pm 0.12$~keV, accompanying the
neutral fluorescent iron line of large equivalent width, $3.4\pm 0.7$~keV
(Table~1, Model 4).  Thus while the Si and S line energies and strengths
are consistent (though with fairly large error bars) with the
fluorescent  lines expected from reflection from X--ray illuminated
neutral material with Solar abundances, the iron line is relatively strong,
which indicates super-Solar abundances in this element.  

\begin{figure}
\epsscale{0.7}
\plotone{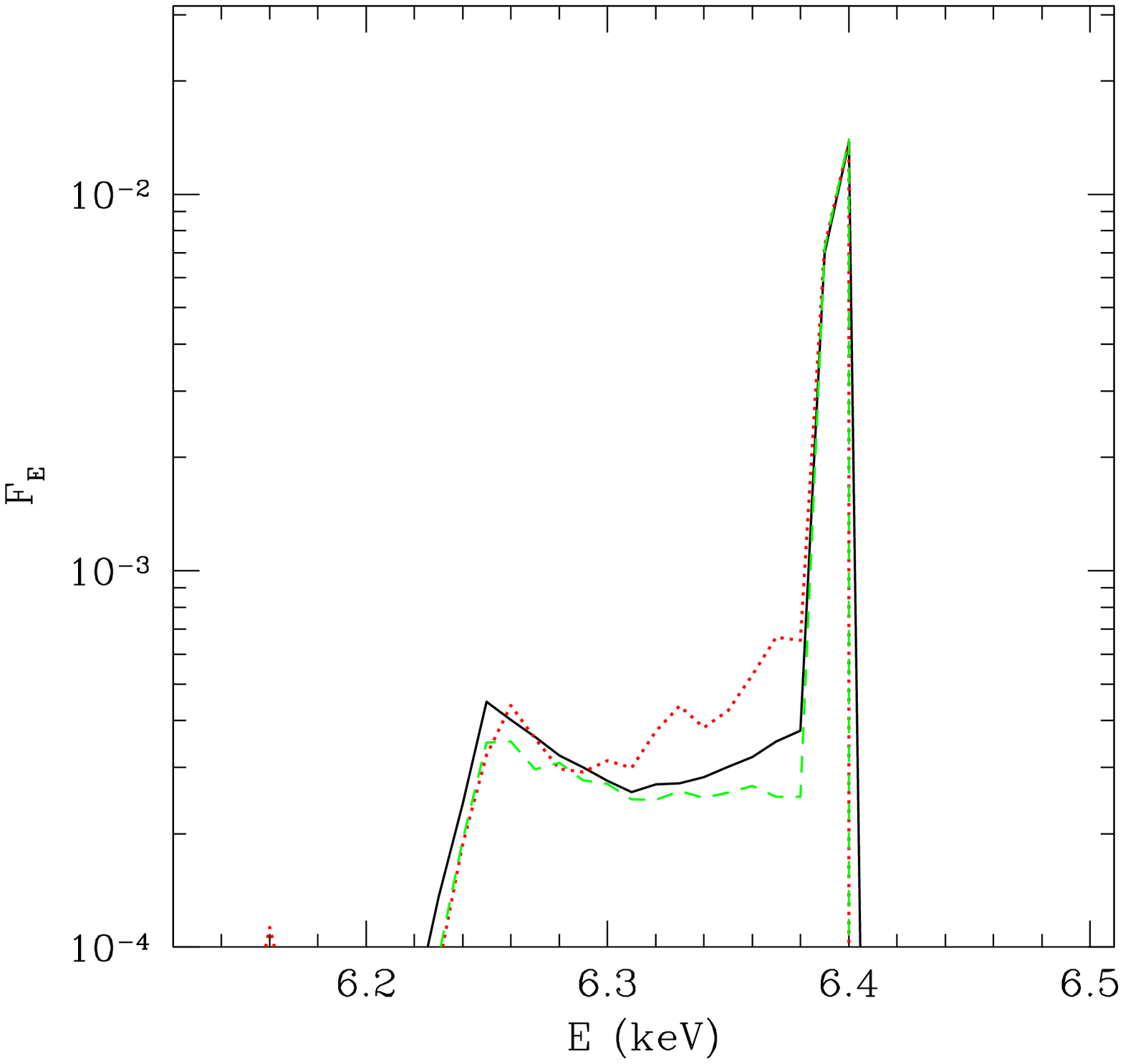}
\caption{Profiles of the Fe K$\alpha$ line
in different geometries, showing
differences in the strength of the Compton shoulder. Solid (black)
curve shows the profile from transmission through a vertically-extended 
structure (assumed to have a geometry of a torus), seen close
to equatorial plane, dominated by reflection from its opposite side.  
Red (dotted) is the profile for pure transmission through the
``torus'' in equatorial plane (opposite side obscured). Green (dashed)
is the profile for usual Compton reflection seen at 30 degrees. Profiles
have been normalized to equal fluxes in the line core.}
\end{figure}

There is some evidence for additional cold photoelectric
absorption, with the best fit yielding $N_H\sim 50-70\times 10^{20}$
cm$^{-2}$ in addition to the $6.5 \times 10^{20}$
cm$^{-2}$ attributable to our Galaxy.  This inference depends on the
description of the hot gas component:  this is because
the abundances and relative
intensities in the {\tt mekal} plasma are strongly correlated with
the inferred level of photoelectric absorption in the spectral fitting.
In any case, such columns are generally
seen towards starbursts, as expected from the molecular gas in these
regions (e.g., Pietsch et al.\ 2001).
Note that this is in contrast to the {\sl primary}
continuum of the AGN which we assume to have {\sl very large}
absorption, $> 10^{24}$ cm$^{-2}$. The model including such
additional absorption is now given in
Table~1 as Model 5.

Although the observed soft X-ray emission is consistent with arising
in a  gas-rich, dusty starburst region, there is a potential
alternative origin via {\em scattering} in partially ionized
material.  Such extended emission from photo-ionized gas is seen in the
soft X-ray spectra of other AGN (e.g.\ Mkn 3; Sako et al.\ 2000).
Here we investigate the origin of this emission by replacing the two
{\tt mekal} plasma components with the emission expected from 
photo-ionized
gas (modeled using the {\sc XSTAR} code: Bautista \& Kallman 2001;
grid19c from the {\sc XSTAR} web page). A simple model, with one {\sc XSTAR}
component replacing the two {\tt mekal} components (from Model 4
in Table~1) gives a bad description of data, $\chi^2_{\nu} = 152/49$ 
(Model 1 in Table~\ref{tab:xstar}). The fit can be improved
by adding a power law representing the primary emission scattered off a
completely ionized plasma, but the fit is still unsatisfactory
($\chi^2_{\nu} = 106/48$;  Model 2 in
Table~\ref{tab:xstar}), with strong residuals around 0.6 and 1 keV.
However, previous fits indicate that the emitting plasma
is multi-temperature, therefore we also try more complex models.
Two photo-ionized {\sc XSTAR} components (with different ionization
parameters) give a better fit, $\chi^2_{\nu}=96.4/47$ (Model 4),
but strong residuals still remain at low energies.  These residuals cannot 
be explained by any pair of photo-ionized components, but an inclusion
of an additional {\tt mekal} component provides a satisfactory fit,
giving $\chi^2_{\nu} = 58.1/45$ (Model 5).  For completeness, we have
also investigated that a simpler hybrid model
with one {\sc XSTAR} and one {\tt mekal} component (Model 3), but this
gives an unsatisfactory fit, $\chi^2_{\nu}=89.4/47$. Thus, at least
some contribution from mechanically-heated plasma is necessary to explain
the soft X-ray emission, but even then, the photo-ionized plasma has to 
have a range of ionization stages.  Presence of mechanically (collisionally)
heated plasma is in fact expected in IC~2560, as the galaxy contains
a starburst region (Cid Fernandes et al. 2004).

To quantify this, we investigated the relative fluxes of the {\tt mekal}
and photo-ionized plasmas.  The two {\tt mekal} components
in Model 5 of Table 1 contribute comparably -- at the observed flux level 
of $\sim 2 \times 10^{-14}$ erg cm$^{-2}$ s$^{-1}$.  In the hybrid model,
Model 5 of Table~2, the {\tt mekal} component contributes comparably
to that in low $T$ component in the two-{\tt mekal} model
($\sim 2 \times 10^{-14}$ erg cm$^{-2}$ s$^{-1}$), while
the sum of the two {\sc XSTAR} components
contributes $\sim 5 \times 10^{-14}$ erg cm$^{-2}$ s$^{-1}$.  This
leads us to the conclusion that the total contribution of photo-ionized
emission to the soft X-ray flux of IC~2560 can be as little as 0\%
and much as $\sim 70$\% of the total, but (1) at least some
collisionally heated plasma is required, and (2) the photo-ionized
component implies a wide range of ionization parameters.

Hence we use the best fit starburst/reflection dominated AGN model
(that given in Table~1 as Model 5) to plot
the unfolded spectrum shown in Fig. 3. A remaining small residual
can be seen at $\sim 5$~keV.  Including a free Gaussian line to model
this feature gives a better fit with $\chi^2_\nu= 46.9/39$, formally a
significant detection with line energy $5.1\pm 0.2$~keV, width
$0.5_{-0.4}^{+0.3}$~keV and very large EW of $1.5_{-1.2}^{+2.0}$~keV.
However, no plausible atomic features exist at this energy.  At
such high equivalent width it is most likely to be iron but the energy
shift is very large, requiring the material to be flowing away from
the observer at $\sim 50,000$ km s$^{-1}$;  a multiple Compton
downscattered shoulder explanation seems even more contrived.
However, this best fit also requires that the reflected emission is
now due to an unusually steep power law illuminating the reflector,
with energy spectral index $\alpha=2.1$.
Assuming that the intrinsic AGN spectrum has a more usual $\alpha=1$
gives $\chi^2_\nu=54.3/40$, compared to $57.7/43$ without the
additional line, so it is not significant on an F-test. Thus it seems
most likely that this feature  may be an artifact in a complex, low
signal-to-noise spectrum.

With such a model, and including the emission lines as given below, the
measured total 2--10 keV flux is $3.3\times 10^{-13}$ erg cm$^{-2}$ 
s$^{-1}$.
When the emission lines discussed below are accounted for separately,
then measured reflection model flux (and this of course does
not include the primary continuum) is $2.6\times 10^{-13}$
erg cm$^{-2}$ s$^{-1}$.  In the extreme case, if the reflector
were to subtend a solid angle as large as $2\pi$, it would have to be
illuminated by an intrinsic, unabsorbed -- but invisible directly to the
observer -- 2--10 keV flux of $4.5\times 10^{-12}$ erg cm$^{-2}$
s$^{-1}$.  A more likely geometry might be one in which much smaller
fraction of the reflector is visible to us, in which case the unabsorbed
flux of the AGN is correspondingly higher.  We can thus set a lower
limit on the 2 -- 10 keV luminosity of $\sim 3 \times 10^{41}$ erg 
s$^{-1}$:
at least this much is needed to account for the reflected spectrum.

\begin{figure}
\epsscale{0.7}
\plotone{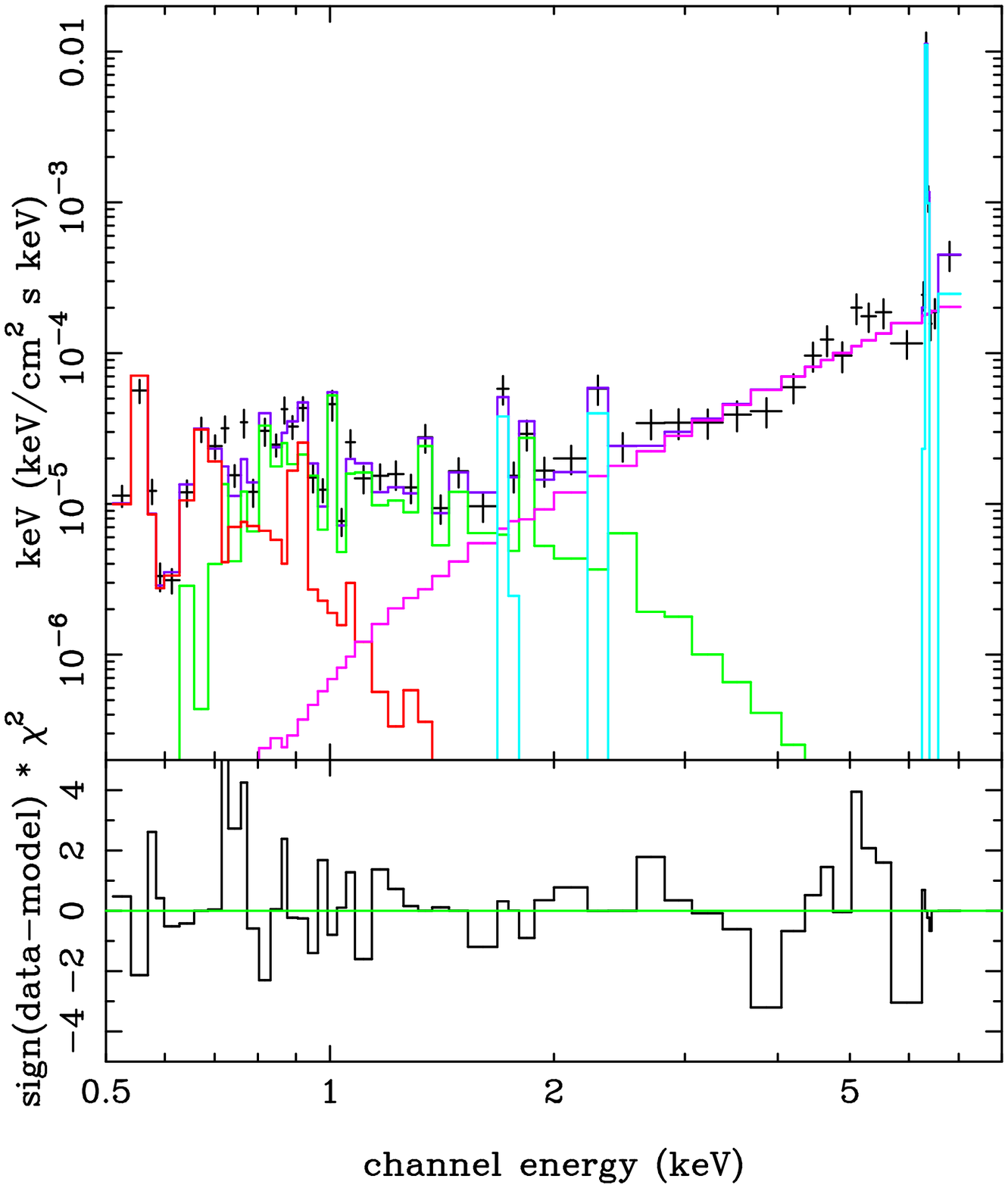}
\caption{Unfolded spectrum of IC~2560, as collected 
by Chandra in 2004.  The model spectrum includes the Compton reflection of 
the invisible continuum, the two thermal plasma components, and the 
emission lines as discussed in the text.}
\end{figure}

\section{Origin of the X-ray spectral features and the relationship 
of the megamaser emission to the X-ray absorption}

The spectral fits presented in the previous section
indicate that  the iron K$\alpha$ line is
somewhat stronger than predicted, perhaps indicating super-Solar iron
abundance. We quantify this by a Monte Carlo simulation of reflection
from a square cross-section torus-like structure with Thomson optical depth
$\tau_T=3$, with inner edge starting $50^\circ$ from the ``torus'' axis
(see Krolik, Madau, \& \.{Z}ycki 1994). This gives a predicted iron
K$\alpha$ line equivalent width of 1.5 keV (Anders-Grevesse
abundances: Fe/H = $4.7\times 10^{-5}$), or 1.2 keV (Anders-Ebihara
abundances: Fe/H = $3.3\times 10^{-5}$), for viewing angles close to
the equatorial plane. To get a reflected iron line 
as large as that observed requires at least 2$\times$ Solar
abundance of Fe. The large iron K$\alpha$ line also should be accompanied
by strong  K$\beta$ emission at $7.05$~keV, 
whose strength is about 10\% of that of the 
K$\alpha$ line. While the signal-to-noise ratio at these high energies
is not high, such a feature (with all parameters fixed or tied to
K$\alpha$) reduces $\chi^2_\nu$ by 3 to $57.8/42$. 

An alternative explanation for a large iron line equivalent width
is the contribution from a heavily absorbed {\em transmitted\/}
spectrum as well as reflection (Ghisellini, Haardt, \& Matt 1994;
Levenson et al.\ 2002).  If the observed line is indeed due to transmission, 
it would imply that the direct component, expected at 20-100~keV, 
should be quite strong, as in the case of NGC~4945 (Done, Madejski, \&
Smith  1996). However, with Solar abundances, this would require the
model torus to subtend a rather large solid angle with a polar
opening angle of $\le \pm15^\circ$ (Levenson et al.\ 2002).  The Compton
downscattered shoulder would also be stronger than that seen in
reflected emission, up to 40\% of the strength of the narrow line core
(Matt 2002;  see also Fig. 2), which is inconsistent with the 15\%
limit on the downscattered  continuum estimated here.

From the spectral fits described in Tables 1 and 2, a picture
emerges where the heavily absorbed primary continuum is reprocessed by
nearly neutral material producing a Compton reflection component as
well as fluorescence lines from Fe, S and Si.  IC~2560 joins the rather
small number of AGN (Mkn~3: Sako et al.\ 2000 and NGC~6552: Reynolds et
al.\  1994) in which the fluorescence lines of species with low
atomic numbers (Si and S) are seen, and thus, reflected continuum in
the 1--10~keV band dominates. The reflector is likely to be formed by
the same optically thick material that obscures a direct view of
the central engine.  In IC~2560, H$_2$O megamaser emission is
also believed to arise from a close to edge-on structure (Ishihara et al.
2001), which by analogy to the well studied case of NGC~4258, is
probably a parsec or sub-parsec radius annulus in an accretion disk.  
Geometrically, the simplest explanation of the reflector, absorber, and
maser are that all three correspond to the same structure, the accretion
disk. (See Greenhill et al. 2003 for discussion of similar
circumstances in Circinus Galaxy.)  We expect to see in reflection only a small
``sliver'' of this X-ray illuminated material on the far side of the
disc (a feature readily accommodated if the disk is warped), as has been
observed in the megamaser galaxies NGC~4258 (Miyoshi et al. 1995) and
Circinus Galaxy (Greenhill et al. 2003). Alternatively, reflection  could
originate from low covering fraction, dense clouds above and below the
obscuring maser structure that blocks  our view of the
central engine. The former explanation has the advantage of being
simpler:  only one high column structure required, while
the latter requires different regions of optically thick material. 

Regardless of the detailed location and geometry of the 
Compton reflector, the covering fraction of reflecting material is
less than unity. This provides a lower limit to the
intrinsic X-ray flux, $\sim 4 \times 10^{-12}$ erg cm$^{-2}$
s$^{-1}$ in the 2--10~keV band, corresponding to a {\sl minimum\/}
bolometric (0.01 - 100 keV) flux of $6 \times 10^{-11}$ erg
cm$^{-2}$ s$^{-1}$ and luminosity
$L = 5 \times 10^{42}$ erg s$^{-1}$.  Adopting the central engine
mass estimated by Ishihara et al. (2001), we can calculate 
$L/L_{\rm Edd} > 0.01$.  If we assume that all of the
observed IR emission ($L = 3 \times 10^{43}$ erg
s$^{-1}$) is reprocessed nuclear flux, then 
a very conservative upper limit is
$L/L_{\rm Edd} < 0.1$. In reality this upper limit could 
be reduced to account for the IR flux that might 
arise in the
circum-nuclear starburst region. With these constraints, we can
infer that the accretion rate is more like that
inferred for the common,  broad-line Seyfert 1 galaxies (e.g.\
Czerny, R\'{o}\.{z}a\'{n}ska, \&  Kuraszkiewicz 2004) and the heavily 
obscured (Seyfert 2) megamaser galaxies NGC~3079 (Kondratko et al. 
2005b), NGC~4945 (Greenhill et al. 1997), and NGC~1068 (Greenhill \& Gwinn 
1997), than it is for
the low $L/L_{\rm Edd}$ core of the
classic megamaser galaxy NGC~4258 (e.g., Lasota et al.\ 1996).  

IC~2560 bears a startling resemblance to NGC~4945 (Done et al.\ 2003;
Schurch, Roberts, \& Warwick 2002) in showing a reflection dominated hard
X-ray spectrum, together with starburst dominated soft X-ray emission. 
In IC~2560, there is evidence for a young ($\le 25$ MYr) stellar
population (Cid Fernandes et al.\ 2004), so it seems likely
that there is hot gas from a  starburst providing the majority of the
soft X-ray flux, rather than photo-ionization. These
characteristics differ from Mkn~3 (Sako et al.\ 2000) and NGC~1068
(Ogle et al.\ 2003), which have clear signatures of predominantly
photo-ionized soft X-ray emission. In both cases, detection of
polarized broad optical emission lines, indicates that there is a
''mirror'' directly reflecting UV and soft X-rays from the central source.
Neither IC~2560 nor NGC~4945 show broad polarized emission
lines.  It is worth noting here that both IC~2560 and NGC~4945 are
powered by relatively low-mass black holes and have modest nuclear
luminosities.  Low intensity in the photo-ionized component
might be simply due to low luminosity of the central source.  

A plausible scenario for many megamaser galaxies (as well as other
Seyfert 2s) is that we view the primary power law continuum of
$\alpha \sim 1$ through an absorber of appreciable optical thickness.  
This can provide an absorbing column due to the neutral gas that might be 
relatively small ($10^{23}$ cm$^{-2}$), where the primary continuum 
shows a modest photoelectric absorption cutoff as in Mkn~3 or NGC~4258.  
Alternatively, the absorber can provide a column of several times
$10^{24}$ cm$^{-2}$ or more, as seen in NGC~4945, NGC~1068, and
NGC~3079 -- and IC~2560 belongs to the latter group (see Table 3).  
For material
that is optically thick to electron scattering, the primary
continuum is depressed at all energies.  For a column with Thomson
optical depth of order unity  ($N_H=3-10\times 10^{24}$ cm$^{-2}$), 
the attenuated  primary spectrum can be seen above 10~keV (as in
NGC~4945), but  for higher columns it is completely suppressed (as in
NGC~1068).  The  column in IC~2560 can only be constrained by the current
Chandra data to be  greater than or equal to that seen in NGC~4945, $N_H
> 3\times 10^{24}$  cm$^{-2}$. Sensitive observations at higher X-ray
energies  are required in order to determine whether the primary emission
can be detected above 10~keV or whether like NGC~1068, IC~2560
 is completely obscured along the line of sight. 

The possibility of a physical association between megamaser
emission and obscuring material raises the question of whether there may
be a correlation between the occurrence of visible megamaser emission and
detection of large obscuring columns. Most importantly, 
maser emission is anisotropic, and for a thin, moderately 
warped disk, it would be beamed in a narrow solid angle about the 
tangent planes (contingent on favorably small gradients in line-of-sight 
velocity).  As well, it is conceivable that maser amplification could be
enhanced (i.e., longer gain paths 
or narrower beam angle) when large columns are
present.  Direct evidence that megamaser emission originates from
disk material is available for about a third of the $\sim 60$ known
masers, chiefly in the form of VLBI maps or spectra that display highly 
Doppler shifted line complexes symmetrically bracketing the systemic
velocity ($V_{sys})$.  
Of these, there are estimates of X-ray absorption columns for
11 galaxies.  Most of these systems are Compton thick; eight
exhibit absorption in excess of 10$^{24}$~cm$^{-2}$ (Table~3).  
Overall, the distribution of column densities
is substantially skewed with respect to the absorption 
distribution for type-2 AGN in general (Bassani et al. 1999; Risaliti et
al. 1999).  However, three known Compton thin cases are notable: NGC~4258,
NGC~4388, and NGC~4051.   Each exhibits variation in column density
(Fruscione et al. 2005; Elvis et al. 2004; McHardy et al. 1995).  In 
the case of NGC~4258, the accretion disk has been shown to cross the line of 
sight to the central source and modest variability in
$N_H$ over months is believed to arise from clumpiness at radii 
comparable to that of the persistent maser emission, on the order 
of $10^4$-$10^5$ Schwarzschild radii 
(Fruscione et al. 2005; Herrnstein et al. 2005).
In contrast, NGC~4388 has displayed extreme variability in column 
density, on time scales as short as hours, and to values as low as 
$2\times10^{21}$ cm$^{-2}$. The inferred radius of the absorber is
$\la 100$~R$_{Sch}$ (Elvis et al. 2004).  
The angular structure and persistence 
of the maser emission is not known, but 
comparison to that of the absorber would enable 
testing of whether both arise from the same (disk-like) structure.  
The third AGN, NGC~4051 is unusual in that it is a narrow line Seyfert 
1 (NLSy1) yet hosts a maser source that has been suggested to 
arise in a central accretion disk 
(Hagiwara et al. 2003).  Column density variation 
is evident (McHardy et al. 1995), but unlike NGC~4258 and NGC~4388, the 
absorber is ionized and unlikely to lie physically close to the maser 
medium.  It is unlikely that there is any edge-on 
structure comprising cold material that could support maser emission, 
and the characteristics of the
maser spectrum is consistent with origin in a wind rather than a disk
(see Table 3 note). This is notable given the importance of nuclear winds
in the formation of NLSy1 spectra. On the other hand, the absorber 
and maser could both trace different parts of a warped structure that 
is only moderately inclined at small radii and tangent to some lines 
of sight at large radii.  A clumpy or frothy medium might also enable 
a line of sight to the central source and ionized absorber at the same 
time as long gain paths are achieved among clumps at larger radii.

\begin{deluxetable}{lc|ccc|c}
\label{tab:highcolumnmasers}
\tablewidth{0pc}
\tablecolumns{6}
\tablecaption{Disk masers in AGN with estimated column densities$^{(1)}$}
\tablehead{
\colhead{Galaxy} & \multicolumn{2}{c}{ID \tablenotemark{(2)}} &
\colhead{$N_H$} &
\multicolumn{2}{c}{References} \\
\colhead{}       & \colhead{VLBI} & \colhead{Spectra} &
\colhead{($10^{23}$ cm$^{-2}$)}  & \colhead{$N_H$} & \colhead{Maser}
}
\startdata

NGC~4945 &  $\surd$ &         & $45\pm4$         & 1 & 13 \\

Circinus Galaxy &  $\surd$ & $\surd$ & $43^{+4}_{-7}$   & 2 & 14\\

M~51     &          & $\surd$ & $58^{+38}_{-18}$ & 3 & 15\\

NGC~1386 &  $\surd$ & $\surd$ & $>22$            & 4 & 16, 24 \\

NGC~1068 &  $\surd$ & $\surd$ & $\ga100$         & 5 & 17 \\

NGC~3079 &  $\surd$ & $\surd$ & $\ga100$         & 6 & 18 \\

IC~2560  &          & $\surd$ & $\ga30$          & 7, 8 & 19 \\

NGC~3393 &          & $\surd$ & $44^{+25}_{-11}$ & 9    & 20 \\

NGC~4258 &  $\surd$ & $\surd$ & $0.6-1.3$   & 10   & 21, 22 \\

NGC~4388 &          & $\surd$ & $0.02-4.8$  & 11  & 23 \\

NGC~4051 &          & $\surd$ & 0 (cold); $0.8-3.7$ (ionized) & 12 & 24 \\

\enddata

\parbox{6.8in}{
$^{(1)}$-- Henkel et al. (2005) and Kondratko et al. (2005b,
2005c) list known H$_2$O masers in AGN. Disk masers are those
originating within accretion disks, at parsec or sub-parsec radii.\\
$^{(2)}$-- Nature of evidence supporting disk
maser identification.  VLBI mapping provides direct evidence. 
Spectra exhibiting red and blueshifted emission (more or less)
symmetrically bracketing $V_{sys}$ provide indirect evidence. For
inclusion here, offsets from systemic must be $>\pm100$ km\,s$^{-1}$.

\smallskip
{\it Notes on specific objects:} \\
{\em NGC~4945}-- Disk-like structure detected, similar to NGC~3079.\\
{\em Circinus Galaxy}-- Column $>10^{25}$ cm$^{-2}$ is admitted in an
alternate model. \\
{\em M~51}-- Red and blueshifted maser emission are intermittent separately.\\
{\em NGC~1386}-- VLBI detection of red side of the disk.  Later
detection of the blue side in spectra. \\ 
{\em NGC~4388}--Classified as a
Seyfert 2 galaxy, but recently,  exhibited large, rapid change in
$N_H$. \\ 
{\em NGC~4051}-- Narrow-line Seyfert 1;  no neutral absorber;   ionized
absorber with significant  variability.  Spectrum comprises narrow
lines distributed over $\sim 280$ km\,s$^{-1}$, though  not well
centered  on $V_{sys}$.   Hagiwara et al. (2003) infer
association with an accretion disk.  However, broadly spread narrow lines
are also consistent with emission from a wind, as in
Circinus (Greenhill et al. 2003). 

\smallskip
{\em References--}
(1) Madejski et al. (2000);
(2) Matt et al. (1999);
(3) Fukazawa et al. (2001);
(4) Guainazzi et al. (2005);
(5) Matt et al. (1997);
(6) Iyomoto et al. (2001);
(7) Iwasawa et al. (2002);
(8) this paper;
(9) Guainazzi et al. (2005);
(10) Fruscione et al. (2005);
(11) Elvis et al. (2004);
(12) McHardy et al. (1995);
(13) Greenhill et al. (1997);  
(14) Greenhill et al. (2003);
(15) Hagiwara et al. (2001);
(16) Braatz et al. (1997);
(17) Greenhill \& Gwinn (1997);
(18) Kondratko et al. (2005b);
(19) Ishihara et al. (2001);
(20) Kondratko et al. (2005c);
(21) Miyoshi et al. (1995);
(22) Greenhill et al. (1995);
(23) Braatz et al. (2004);
(24) Hagiwara et al. (2003) 
}
\end{deluxetable}

\section{Summary and conclusions}  

Chandra observation of the H$_2$O megamaser galaxy IC~2560
reveals a complex X-ray spectrum.  At low energies, it exhibits a 
soft spectral component that is unlikely to be
characterized by a single-temperature, collisionally-ionized plasma, 
as this would require 
anomalously low elemental abundances.  Instead, the soft component
probably arises from such plasma -- presumably associated with the host galaxy
-- at a range of temperatures, 0.1 keV $< kT <$ 0.7 keV, with
abundances close to Solar.  It is also possible to fit the spectrum by  a
model that combines one collisionally ionized 
plasma component and a complex photo-ionized 
scattering medium, with a range of ionization parameters.  

At higher energies, the spectrum is dominated by a very hard X-ray
continuum, together with several emission line features  at energies
consistent with neutral Fe, Si, and S 
K$\alpha$ lines.  The best interpretation of such a 
hard spectrum is that the primary continuum is entirely absorbed (at
least in the Chandra bandpass), requiring $N_H >0.3 \times 10^{25}$
cm$^{-2}$, and that we observe only the much harder Compton 
reflection of the hidden primary continuum from optically thick, 
high column density material.  Given that the H$_2$O megamaser emission
is believed to arise in a close to edge-on disk-like structure and that
maser action requires a large column of material, the simplest 
interpretation is that all three high column, low-ionization  structures
(absorber, reflector and megamaser) arise from a single physical
component, probably a pc-scale or smaller accretion disk.  
We hypothesize that this is the case more generally and anticipate 
a correlation of megamaser emission and high X-ray obscuration.  In an
initial sample of 11 galaxies, this hypothesis is borne out.  
Discovery of new  maser sources, identification of emission
origins for known sources, and measurement of X-ray spectra for host AGN
will enable further testing.  

The Compton reflection from this material also produces strong 
fluorescent emission lines.  The observed intensities
of the Si and S lines are entirely consistent with reprocessing 
by Solar abundance material, while the Fe K line, with equivalent
width  of $\sim 2.5$ keV, is a factor of 2 higher than expected.    
The line might arise in reflection, transmission, or both.  
However, the most likely explanation for the high  equivalent width
of the line is  Compton reflection from neutral medium with
super-Solar ($\sim 2 \times$)  Fe abundances.  This is further supported 
by the estimated upper limit on the intensity of the Compton
``shoulder''  that should be associated with the Fe K line.  

The intensity of the reprocessed component provides 
a lower limit on the 2 - 10 keV flux of the unabsorbed primary continuum 
of $4\times 10^{-12}$ erg cm$^{-2}$ s$^{-1}$, corresponding to 
a lower limit on the 0.01 - 100 keV luminosity of $\sim 5 \times 10^{42}$ 
erg s$^{-1}$.  Conversely, the observed bolometric luminosity of
$\sim 3 \times 10^{43}$ ers s$^{-1}$ places an upper limit. 
Since the mass of the central source estimated from partial
resolution of the megamaser angular structure is $\sim 3 \times
10^{6}$ M$_\odot$, we infer that the source
is accreting at a moderate rate corresponding to 
$0.01< L/L_{Edd} < 0.1$, as is the case for
other heavily absorbed megamaser galaxies (e.g., NGC~1068).  Further
clues to the structure of this  source are likely to be revealed by
sensitive hard X-ray observations  at 10 - 50 keV, with missions such as
NASA's NuSTAR.  

\acknowledgments

This project was partially supported 
by Chandra grants no. GO4-5125X and GO4-5127X from NASA via Smithsonian
Astrophysical Observatory, by the Polish KBN grants 
2P03D01225 and PBZ-KBN-054/P03/2001, 
and by the Department of Energy
contract to SLAC no. DE-AC3-76SF00515. L. G. thanks SLAC as well as KIPAC 
(Stanford University) for the hospitality at the time when this 
research was conducted.


\begin{thebibliography}{}

\bibitem[Bassani et al. 1999]{Bassani99} 
Bassani, L., et al. 1999, ApJS, 212, 473 


\bibitem[Bautista and Kallman 2001]{Kal2001} Bautista, M. A., \& 
Kallman, T. 2001, ApJS, 134, 139

\bibitem [Braatz, Wilson, \& Henkel 1996] {braatz1996} Braatz, J. A., 
Wilson, A. S., \& Henkel, C. 1996, ApJS, 106, 51 

\bibitem[Braatz et al. (1997)]{braatz97} Braatz, J., Greenhill, L., 
Moran, J., Wilson, A., \& Herrnstein, J. 1997, Bull. A. A. S., 29, 1374

\bibitem[Braatz et al. (2004)]{braatz04} Braatz, J. A., Henkel, 
C., Greenhill, L. J., Moran, J. M., \& Wilson, A. S. 2004, ApJ, 617, L29

\bibitem[Cid Fernandes et al. 2004]{cidf04}Cid Fernandes, R., et al. 
2004, ApJ, 605, 127

%\bibitem[Comastri (2004)]{comastri04}Comastri, A 2004,
%astro-ph/0403693

\bibitem[Czerny et al. 2004]{Czerny2004} Czerny, B., R\'{o}\.{z}a\'{n}ska A., 
\& Kuraszkiewicz J. 2004, A\&A, 428, 39

\bibitem [Done et al. 2003] {done2003}  Done, C., Madejski, G. M., \.{Z}ycki, 
P. T., \& Greenhill, L. J. 2003, ApJ, 588, 763

\bibitem [Done et al. 1996]{Done96} Done, C., Madejski, G., \& Smith, D. 
1996, ApJ, 463, L63

\bibitem[Elvis et al.(2004)]{elvis2004} Elvis, M., Risaliti, G., 
Nicastro, F., Miller, J.~M., Fiore, F., \& Puccetti, S.\ 2004, \apjl, 615, 
L25 

\bibitem[Fruscione et al.(2005)]{fruscione2005} Fruscione,
A.,  Greenhill, L.~J., Filippenko, A.~V., Moran, J.~M., Herrnstein,
J.~R., \&  Galle, E.\ 2005, \apj, 624, 103

\bibitem[Fukazawa et al.(2001)]{fuluzawa01} Fukazawa, Y., Iyomoto, 
N., Kubota, A., Matsumoto, Y., \& Makishima, K.\ 2001, A\&A, 374, 73

\bibitem[George and Fabian (1991)] {george1991}
George, I. M., \& Fabian, A. C. 1991, MNRAS, 249, 352

\bibitem[Ghez et al.(2005)]{Ghez2005} Ghez, A.~M., Salim, 
S., Hornstein, S.~D., Tanner, A., Lu, J.~R., Morris, M., Becklin, 
E.~E., \& Duch{\^ e}ne, G.\ 2005, ApJ, 620, 744

\bibitem[Ghisellini, Haardt, and Matt 1994]{Ghis1994} 
Ghisellini, G., Haardt, F., \& Matt, G. 1994, MNRAS, 267, 743

\bibitem[Greenhill et al. 1995]{Green95} Greenhill, L. J., Jiang, D. R., 
Moran, J. M., Reid, M. J., Lo, K. Y., \& Claussen, M. J. 1995, ApJ, 440, 619

\bibitem[Greenhill et al.(1997)]{greenhill97} Greenhill, L.~J.,
Moran, J.~M., \& Herrnstein, J.~R. 1997, ApJL, 481, L23

\bibitem[Greenhill \& Gwinn 1997]{Green97} Greenhill, L. J., \& Gwinn, C.  
1997, Ap\&SS, 248, 261

\bibitem[Greenhill et al. 2003]{Green03} Greenhill, L. J., et al. 
2003, ApJ, 590, 162

\bibitem[Guainazzi et al.(2005)]{guainazzi05} Guainazzi, M., 
Fabian, A.~C., Iwasawa, K., Matt, G., \& Fiore, F.\ 2005, MNRAS, 356,
295

\bibitem[Hagiwara et al. 2003]{Hagiwara2003} Hagiwara, Y., Diamond, P., 
Miyoshi, M., Rovilos, E., \& Baan, W. 2003, MNRAS, 344, L53

\bibitem[Hagiwara et al. (2001)]{hagiwara01} Hagiwara, Y., Henkel, C., 
Menten, K. \& Nakai, N. 2001, ApJ, 560, 37

\bibitem[Henkel et al. 2005]{Henkel05} Henkel, C., et al. 2005, 
Ap\&SS, 295, 107

\bibitem[Illarionov et al. 1979]{Ill1979} Illarionov, A., Kallman, T., 
McCray, R., \& Ross, R. 1979, ApJ, 228, 279

\bibitem[Ishihara, Y., et al. (2001)]{Ish2001} Ishihara, Y., 
Nakai, N., Iyomoto, N., Makishima, K., Diamond, 
P., \& Hall, P. 2001, PASJ, 53, 215

\bibitem[Iwasawa, Maloney, and Fabian (2002)]{IMF02} Iwasawa, K., 
Maloney, P. R., \& Fabian, A. C. 2002, MNRAS, 336, L71

\bibitem[Iyomoto et al.(2001)]{iyomoto01} Iyomoto, N., Fukazawa, 
Y., Nakai, N., \& Ishihara, Y.\ 2001, ApJ, 561, L69 

\bibitem[Kondratko et al. 2005a]{kondr05} Kondratko, P., 
Greenhill, L., \& Moran, J. 2005, ApJ, 618, 681

\bibitem[Kondratko et al. 2005b]{kondratko05b}Kondratko, P. T., et al.
2005, ApJ, submitted

\bibitem[Kondratko et al. 2005c]{kondratko05c} Kondratko, P. T.,
Greenhill, L. J., Moran, J. M. 2005, ApJ, in prep

\bibitem[Krolik, Madau, \& Zycki 1994]{krol94} Krolik, J.,
 Madau, P., \& \.{Z}ycki, P. 1994, ApJ, 420, L57

\bibitem[Kumar 1999]{kumar99}Kumar, P. 1999, ApJ, 519, 599

\bibitem[Lasota et al. 1996]{las96}Lasota, J.-P., Abramowicz, M., 
Chen, X., Krolik, J., Narayan, R., \& Yi, I. 1996, ApJ, 462, 142

\bibitem[Levenson et al. 2002]{Levenson2002} Levenson, N. A., Krolik, J. H., 
\.{Z}ycki, P. T., Heckman, T. M., Weaver, K. A., Awaki, H., \& 
Terashima, Y. 2002, ApJ, 573, L81

\bibitem[Lodato \& Bertin 2004]{Lodato03} Lodato, G., \& Bertin, G. 2003, 
A\&A, 398, 517

\bibitem[Madejski et al.(2000)]{madejski00} Madejski, G., {\. 
Z}ycki, P., Done, C., Valinia, A., Blanco, P., Rothschild, R., \& Turek, 
B.\ 2000, ApJ, 535, L87

\bibitem[Magdziarz and Zdziarski (1995)]{Magdziarz1995} Magdziarz, 
P., \& Zdziarski, A. 1995, MNRAS, 273, 837

\bibitem[Matt, Brandt, and Fabian (1996)]{Matt1996} Matt, G., 
Brandt, W. N., \& Fabian, A. C. 1996, MNRAS, 280, 823

\bibitem[Matt, Fabian, and Reynolds (1997)]{Matt1997} Matt, G., Fabian, 
A. C. \& Reynolds, C. S. 1997, MNRAS, 289, 175

\bibitem[Matt et al.(1997)]{matt97} Matt, G., et al.\ 1997, 
A\&A, 325, L13

\bibitem[Matt et al.(1999)]{matt99} Matt, G., et al.\ 1999, 
A\&A, 341, L39

\bibitem[Matt 2002]{Matt2002} Matt, G. 2002, MNRAS, 337, 147

\bibitem[Matt et al.(2004)]{2004A&A...414..155M} Matt, G., Bianchi, S., 
Guainazzi, M., \& Molendi, S.\ 2004, \aap, 414, 155 

\bibitem[McHardy et al.(1995)]{mchardy95} McHardy, I.~M., Green, 
A.~R., Done, C., Puchnarewicz, E.~M., Mason, K.~O., Branduardi-Raymont, G., 
\& Jones, M.~H.\ 1995, \mnras, 273, 549

\bibitem[Molendi et al.(2003)]{2003MNRAS.343L...1M} Molendi, S., Bianchi, 
S., \& Matt, G.\ 2003, \mnras, 343, L1 

\bibitem[Myioshi, M., et al. (1995)]{Myoshi1995} Miyoshi, M., 
Moran, J., Herrnstein, J., Greenhill, L., Nakai, N., Diamond, P. \& 
Inoue, M. 1995, Nature, 373, 127

\bibitem[Neufeld, D. et al. 1994]{Neu94}Neufeld, D., 
Maloney, P., \& Conger, S. 1994, ApJ, 436, L127

\bibitem[Ogle et al. 2003]{ogle03} Ogle, P. M., Brookings, T., 
Canizares, C. R., Lee, J. C., \&  Marshall, H. L. 2003, A\&A, 402, 849

\bibitem[Pietsch et al.(2001)]{2001A&A...365L.174P} Pietsch, W., et al.\ 
2001, \aap, 365, L174 

%\bibitem[Pounds et al.(1994)]{1994MNRAS.267..193P} Pounds, K.~A.,
%Nandra,  K., Fink, H.~H., \& Makino, F.\ 1994, \mnras, 267, 193 
 
\bibitem[Reynolds et al. 1994]{rey94} Reynolds, C. S., Fabian, A. C., 
Makishima, K., Fukazawa, Y., \& Tamura, T. 1994, MNRAS, 
268, L55

\bibitem[Risaliti et al. 1999]{Risaliti99} Risaliti, G., Maiolino, R., 
\& Salvati, M. 1999, ApJ, 522, 157

\bibitem[Sako et al. 2000]{Sako2000} Sako, M., Kahn, S., Paerels, F., 
and Liedhal, D. 2000, ApJ, 543, L115

\bibitem[Schurch et al. 2002]{Sch02} Schurch N. J., Roberts T. P., \& 
Warwick R. S. 2002, MNRAS, 335, 241

\bibitem[Sch{\" o}del et al.(2003)]{Schodel2003} Sch{\" o}del, 
R., Ott, T., Genzel, R., Eckart, A., Mouawad, N., \& Alexander, T.\ 
2003, ApJ, 596, 1015

\bibitem[Strickland and Stevens 2000]{Strickland2000} Strickland, D. K., 
\& Stevens, I. 2000, MNRAS, 314, 511

\bibitem[Zycki and Czerny (1994)]{Zycki1994} \.{Z}ycki, P. T., \& Czerny, B. 
1994, MNRAS, 266, 653

\end{thebibliography}
\end{document}